\documentstyle[12pt,axodraw]{article}
\newcommand{\NP}{Nucl. Phys. }
\newcommand{\PR}{Phys. Rev. }
\newcommand{\PRL}{Phys. Rev. Lett. }
\newcommand{\PL}{Phys. Lett. }

\begin{document}
\baselineskip=20pt

\pagenumbering{arabic}

\vspace{1.0cm}
\begin{center}
{\Large\sf Anomalous C-violating three photon decay of the neutral 
pion in noncommutative quantum electrodynamics}\\[10pt]
\vspace{.5 cm}

{Harald Grosse$^a$, Yi Liao$^{b}$}
\vspace{1.0ex}

{\small $a$ Institut f\"ur Theoretische Physik, Universit\"at Wien,\\
Boltzmanngasse 5, A-1090 Wien, Austria\\}

\vspace{1.0ex}

{\small $b$ Institut f\"ur Theoretische Physik, Universit\"at Leipzig,
\\
Augustusplatz 10/11, D-04109 Leipzig, Germany\\}

\vspace{2.0ex}

{\bf Abstract}
\end{center}
We show that a simple and reasonable generalization of the anomalous
interaction between the neutral pion and two photons can induce the
C-violating three photon decay of the neutral pion in noncommutative 
quantum electrodynamics. We find that it is mandatory for consistency
reasons to include simultaneously the normal neutral pion and 
photon interaction in which the neutral pion transforms under $U(1)$ 
in a similar way as in the adjoint representation of a non-Abelian 
gauge theory. 
We demonstrate that the decay has a characteristic distribution
although its rate still seems
too small to be experimentally reachable in the near future. We also
describe how to manipulate phase space integration correctly when
Lorentz invariance is lost. 

\begin{flushleft}


\end{flushleft}

\newpage

Noncommutative field theories have recently received a lot of 
attention mainly because of their connection to string theories 
$\cite{string}$. But Noncommutative field theories are certainly 
interesting in their own right. A possible way to construct the 
noncommutative version of a field theory from its ordinary 
commutative counterpart is by replacing the usual product of fields 
in the action with the $\star$-product of fields. The $\star$-product 
of the two fields $\phi_1(x)$ and $\phi_2(x)$ is defined as 
\begin{equation}
\displaystyle(\phi_1\star\phi_2)(x)=\left[\exp\left(\frac{i}{2}
\theta^{\mu\nu}\partial^x_{\mu}\partial^y_{\nu}\right)
\phi_1(x)\phi_2(y)\right]_{y=x},
\end{equation}
where $\theta^{\mu\nu}$ is a real antisymmetric constant matrix that
parameterizes the noncommutativity of spacetime,
\begin{equation}
\displaystyle[x^{\mu},x^{\nu}]=i\theta^{\mu\nu},
\end{equation}
and has dimensions of length squared.

The noncommutative quantum electrodynamics (NCQED) of photons is then
given by the following Lagrangian $\cite{ncqed}$,
\begin{equation}
\begin{array}{rcl}
{\cal L}_F&=&\displaystyle-\frac{1}{4}F^{\mu\nu}\star F_{\mu\nu},\\
\displaystyle F_{\mu\nu}&=&\partial_{\mu}A_{\nu}-\partial_{\nu}A_{\mu}
+ie[A_{\mu},A_{\nu}]_{\star},
\end{array}
\end{equation}
where the Moyal bracket is defined as
\begin{equation}
\displaystyle
[\phi_1,\phi_2]_{\star}=\phi_1\star\phi_2-\phi_2\star\phi_1.
\end{equation}
The action $\displaystyle\int d^4x~{\cal L}_F$ is invariant under the 
generalized $U(1)$ gauge transformation,
\begin{equation}
\displaystyle
A_{\mu}\to A^{\prime}_{\mu}=U\star A_{\mu}\star U^{-1}+
ie^{-1}U\star\partial_{\mu}U^{-1},~~U(x)=(\exp(i\lambda(x)))_{\star},
\end{equation}
under which $F_{\mu\nu}$ transforms as follows,
\begin{equation}
\displaystyle
F_{\mu\nu}\to F^{\prime}_{\mu\nu}=U\star F_{\mu\nu}\star U^{-1}.
\end{equation}
Note that the neutral photon interacts with itself due to the
Moyal bracket term in $F_{\mu\nu}$ as in the usual non-Abelian gauge
theory. One must then be careful with the gauge fixing procedure 
since the ghost also interacts as opposed to the usual QED in which 
it is free,
\begin{equation}
\displaystyle
{\cal L}_{gf}=-\frac{1}{2\xi}(\partial A)^2+\left(\partial^{\mu}
\bar{c}\right)\left(\partial_{\mu}c+ie[A_{\mu},c]_{\star}\right),
\end{equation}
where we have freely replaced one of the star products by the usual
one because one always has in the action, 
$\displaystyle\int d^4x~\phi_1\star \phi_2=\int d^4x~\phi_1\phi_2$.
The matter fields can also be incorporated. The interested reader 
should consult the above references for details.

The phenomenological implications of the NCQED have began to appear
very recently. They are roughly classified into two categories, 
namely the small corrections to precisely measured quantities in 
low energy atomic systems $\cite{atomic}$ and the relatively larger 
corrections to QED processes at future high energy linear colliders 
$\cite{collider}$. In this work we shall consider a novel combined
effect of the NCQED and the generalized axial anomaly in noncommutative
spacetime, i.e., the three photon ($\gamma$) decay of the neutral pion
($\pi^0$). This decay is a C-violating process which proceeds in the 
standard model through weak interactions and is thus very small. The 
appearance of multi-photons in the final state suppresses further the 
decay because gauge invariance demands many factors of the photon 
momenta whose dimensions have to be balanced by some heavier mass scales.
Indeed, an estimate $\cite{dicus}$ showed its branching ratio in the 
standard model is of order $10^{-31}$, too tiny to be experimentally 
feasible. This makes the decay a possible testing ground for C violation
beyond weak interactions, for example, in electromagnetic or strong
interactions. Our basic observations are two fold. The NCQED violates
the C symmetry $\cite{cpt}$ through the Moyal bracket term and this 
should have some positive effect on the decay $\pi^0\to 3\gamma$ because
the latter may happen at the electromagnetic strength. On
the other hand, since the Lorentz invariance is spoiled by the 
constant matrix $\theta_{\mu\nu}$ we may have completely different 
gauge invariant structures for the decay amplitude so that the 
suppression introduced by gauge invariance may become less severe. If 
both are true we shall expect a large enhancement of the decay.

The neutral pion decays dominantly into two photons, which is driven by
the axial anomaly,
\begin{equation}
\displaystyle{\cal L}_A=
-J\epsilon^{\mu\nu\rho\sigma}\pi^0F_{\mu\nu}F_{\rho\sigma},
\end{equation}
where $F_{\mu\nu}$ is the usual QED electromagnetic tensor, and the 
constant $\displaystyle J=\frac{N_c e^2}{96\pi^2 f_{\pi}}$ with $N_c$ 
being the color number and $f_{\pi}$ the pion decay constant. This 
term is unique in the Wess-Zumino-Witten action $\cite{wzw}$ in the 
sense that it involves the least number of Goldstone bosons while 
involving the most number of photons. Our argument is that this term 
should be extended most naively if there is any kind of generalization
of the axial anomaly to noncommutative spacetime. Then a simple 
and reasonable guess is
\begin{equation}
\displaystyle{\cal L}_{NCA}=
-J\epsilon^{\mu\nu\rho\sigma}\pi^0F_{\mu\nu}\star F_{\rho\sigma},
\end{equation}
where $F_{\mu\nu}$ is now the NCQED electromagnetic tensor appearing
in ${\cal L}_F$ and once again we have dropped the star of $\pi^0$
with $F$'s. Furthermore, the above term should not be affected by
generalization of other terms in the Wess-Zumino-Witten action due to
the same reason of uniqueness. This guess has got some support from
recent one loop approaches to the anomaly in noncommutative spacetime 
$\cite{nca}$.

We notice that ${\cal L}_{NCA}$ contains the desired 
$\pi^0\to 3\gamma$ transition term. So, together with 
${\cal L}_{F}+{\cal L}_{gf}$ we might think we were already prepared to 
calculate the decay amplitude. Actually, for consistency reasons to 
be explained later on, we are still missing one piece: we have to 
include simultaneously the normal direct interactions between the 
photon and $\pi^0$. As shown in Hayakawa's papers in Refs.
$\cite{ncqed}$, besides the possibility of being invariant,
the neutral particle fields may also undergo a
nontrivial transformation under $U(1)$,
\begin{equation}
\displaystyle
\pi^0\to \pi^{0\prime}=U\star\pi^0\star U^{-1}.
\end{equation}
which resembles the adjoint transformation in the usual non-Abelian
gauge theory. Since $F_{\mu\nu}$ itself also transforms under $U(1)$,
it seems that we must include also the above one to keep uniformness
and completeness among neutral fields. The covariant derivative for
$\pi^0$ is
\begin{equation}
\displaystyle
D_{\mu}\pi^0=\partial_{\mu}\pi^0+ie[A_{\mu},\pi^0]_{\star},
\end{equation}
which transforms similarly to $\pi^0$ and becomes trivial in the
usual commutative spacetime. Finally, the NCQED Lagrangian for 
$\pi^0$ can be written down,
\begin{equation}
\displaystyle
{\cal L}_{\pi^0}=\frac{1}{2}D_{\mu}\pi^0\star D^{\mu}\pi^0.
\end{equation}

With all pieces at hand we can now compute the decay amplitude
for $\pi^0(p)\to\gamma(k_1,\epsilon_{1\alpha})+
\gamma(k_2,\epsilon_{2\beta})+\gamma(k_3,\epsilon_{3\gamma})$,
where $p$ and $k_i$ are incoming and outgoing momenta of the $\pi^0$ 
and photons, $\epsilon_i$ are photon polarization vectors with 
Lorentz indices $\alpha,~\beta,~\gamma$. The corresponding Feynman 
diagrams are shown in Fig. 1. Let us first list the relevant 
Feynman rules to set up our notation. The point for the derivation
of them is the recursive use of the following Fourier transformation
for the star product of functions,
\begin{equation}
\displaystyle (\phi_1\star\phi_2)(x)=
\int\frac{d^4k_1}{(2\pi)^4}\int\frac{d^4k_2}{(2\pi)^4}
\tilde{\phi}_1(k_1)\tilde{\phi}_2(k_2)\exp[-i(k_1+k_2)\cdot x]
\exp[-ik_1\theta k_2/2],
\end{equation}
where $\tilde{\phi}_i(k_i)$ are Fourier transforms of $\phi_i(x)$
and $k_1\theta k_2=\theta_{\mu\nu}k_1^{\mu}k_2^{\nu}$.
In the following list of Feynman rules we use $q_i$ to denote the
incoming momenta of photons and ghosts, $p_i$ the incoming
momenta of the $\pi^0$'s and $\mu$ or $\mu_i$ the photon Lorentz
indices. The Feynman rule for the three photon vertex is
\begin{equation}
\displaystyle
+2e~\sin(q_1\theta q_2/2)~V_3^{\mu_1\mu_2\mu_3}(q_1,q_2,q_3),
\end{equation}
with
\begin{equation}
\displaystyle
V_3^{\mu_1\mu_2\mu_3}(q_1,q_2,q_3)=(q_1-q_2)^{\mu_3}g^{\mu_1\mu_2}
+(q_2-q_3)^{\mu_1}g^{\mu_2\mu_3}+(q_3-q_1)^{\mu_2}g^{\mu_3\mu_1}.
\end{equation}
Note that the above vertex actually satisfies the Bose symmetry due 
to momentum conservation and antisymmetry of $\theta_{\mu\nu}$.
The ghost-anti-ghost-photon vertex is
\begin{equation}
\displaystyle
+2e~q_{2\mu}~\sin(q_1\theta q_2/2),
\end{equation}
where $q_2$ is the incoming anti-ghost momentum.
The vertex for the anomalous $\pi^0\to 2\gamma$ transition is 
modified to be
\begin{equation}
\displaystyle -i8J~\epsilon^{\mu_1\mu_2\rho\sigma}q_{1\rho}q_{2\sigma}
\cos(q_1\theta q_2/2).
\end{equation}
The new vertex for the contact $\pi^0 3\gamma$ interaction will
appear as part of the contribution to the decay amplitude and will
be presented below. The final piece of Feynman rule for the 
$\gamma 2\pi^0$ vertex is
\begin{equation}
\displaystyle -2e~(p_1-p_2)_{\mu}\sin(p_1\theta p_2/2),
\end{equation}
which again satisfies the Bose symmetry with respect to $\pi^0$'s.

The separate contributions to the decay amplitude from Fig. 1(a)-(c) 
can be cast in the form,
\begin{equation}
\begin{array}{rcl}
{\cal A}&=&\displaystyle i16eJ\left({\cal A}_a^{\alpha\beta\gamma}+
{\cal A}_b^{\alpha\beta\gamma}+{\cal A}_c^{\alpha\beta\gamma}\right)
\epsilon^*_{1\alpha}\epsilon^*_{2\beta}\epsilon^*_{3\gamma},\\
{\cal A}_a^{\alpha\beta\gamma}&=&\displaystyle 
\epsilon^{\mu\alpha\beta\gamma}\left(A_1k_{1\mu}+A_2k_{2\mu}+
A_3k_{3\mu}\right),\\
{\cal A}_b^{\alpha\beta\gamma}&=&\displaystyle 
\frac{A_1}{2d_1}\epsilon_{\rho}^{~\alpha\mu\nu}(k_2+k_3)_{\mu}k_{1\nu}
V_3^{\beta\gamma\rho}(k_2,k_3,-k_2-k_3)\\
&+&\displaystyle 
\frac{A_2}{2d_2}\epsilon_{\rho}^{~\beta\mu\nu}(k_3+k_1)_{\mu}k_{2\nu}
V_3^{\gamma\alpha\rho}(k_3,k_1,-k_3-k_1)\\
&+&\displaystyle 
\frac{A_3}{2d_3}\epsilon_{\rho}^{~\gamma\mu\nu}(k_1+k_2)_{\mu}k_{3\nu}
V_3^{\alpha\beta\rho}(k_1,k_2,-k_1-k_2),\\
{\cal A}_c^{\alpha\beta\gamma}&=&\displaystyle 
\frac{A_1-A_2}{2(d_1+d_2)}\epsilon^{\alpha\beta\mu\nu}k_{1\mu}
k_{2\nu}\left(k_3+2(k_1+k_2)\right)^{\gamma}\\
&+&\displaystyle 
\frac{A_2-A_3}{2(d_2+d_3)}\epsilon^{\beta\gamma\mu\nu}k_{2\mu}
k_{3\nu}\left(k_1+2(k_2+k_3)\right)^{\alpha}\\
&+&\displaystyle 
\frac{A_3-A_1}{2(d_3+d_1)}\epsilon^{\gamma\alpha\mu\nu}k_{3\mu}
k_{1\nu}\left(k_2+2(k_3+k_1)\right)^{\beta},\\
\end{array}
\end{equation}
where $d_i=k_j\cdot k_k$ with $(i,j,k)$ being cyclic of $(1,2,3)$.
With three independent momenta $k_i$ we can form three independent 
angles together with $\theta_{\mu\nu}$, $\alpha_i=k_j\theta k_k/2$.
Then, $A_i=\sin(\alpha_i)\cos(\alpha_j-\alpha_k)$. Note the Bose
symmetry is separately satisfied by the three contributions. However
the above amplitude does not satisfy the usual QED Ward identity,
for example, 
\begin{equation}
\left({\cal A}_a^{\alpha\beta\gamma}+
{\cal A}_b^{\alpha\beta\gamma}+{\cal A}_c^{\alpha\beta\gamma}\right)
k_{1\alpha}\neq 0;
\end{equation}
instead, we have,
\begin{equation}
\left({\cal A}_a^{\alpha\beta\gamma}+
{\cal A}_b^{\alpha\beta\gamma}+{\cal A}_c^{\alpha\beta\gamma}\right)
k_{1\alpha}k_{2\beta}k_{3\gamma}= 0,
\end{equation}
which is consistent with the effective non-Abelian nature of 
$F_{\mu\nu}$. This also implies that we must be careful when 
computing the physical amplitude squared and summed over polarization
states. As we do with the gluons in the usual QCD, we have basically
two ways to do so. We may use the unphysical polarization sums for
the photons,
\begin{equation}
\displaystyle\sum_{\rm pol}\epsilon^*_{\mu}\epsilon_{\nu}
=-g_{\mu\nu},
\end{equation}
and then remove the unphysical polarization contribution by
subtracting off the ghost contribution to the amplitude squared.
The ghost amplitude for the diagram shown in Fig. 1(d) is
\begin{equation}
{\cal A}_d=+i16eJ~\frac{A_3}{2d_3}\epsilon^{\gamma\mu\nu\rho}
k_{1\mu}k_{2\nu}k_{3\rho}\epsilon^*_{3\gamma}.
\end{equation}
The contribution to be subtracted off, including all cases similar
to Fig. 1(d), is,
\begin{equation}
2(16eJ)^2~\left[(A_1)^2\frac{d_2d_3}{2d_1}+
(A_2)^2\frac{d_3d_1}{2d_2}+(A_3)^2\frac{d_1d_2}{2d_3}\right],
\end{equation}
where the factor $2$ accounts for the interchange of ghost and 
anti-ghost.
The second way is that we use physical polarization sums for the
photons so that only the physical polarization contributions are
kept in the amplitude squared. A convenient form is,
\begin{equation}
\displaystyle\sum_{\rm pol}\epsilon_{i\mu}^*\epsilon_{i\nu}
=-g_{\mu\nu}+(k_{i\mu}n_{i\nu}+k_{i\nu}n_{i\mu})(k_i\cdot n_i)^{-1},
\end{equation}
where $n_i$ is an arbitrary vector satisfying $n_i^2=0$ and 
$k_i\cdot n_i\neq 0$. In practice it is most convenient to choose 
$n_i$ as any of the other two photon momenta. 

Now the physical result should not depend on the ways of how to 
incorporate physical polarizations. But we found that without 
considering the contribution in Fig. 1(c) orginating from the normal 
$\pi^0$-photon interaction ${\cal L}_{\pi^0}$ there is no way to 
achieve an identical result in the above two ways. However, 
including that contribution leads to a unique result which is 
independent of the ways to do polarization sums and especially 
of how to choose $n_i$ for the $i$-th photon. This indicates 
unambiguously that for consistency of the calculation the 
$\pi^0$ field must transform under $U(1)$ nontrivially as shown
in eq. (10) instead of being invariant and that neutral particles 
must interact with photons in NCQED.

We are now in a position to compute the decay rate. To make this
easier to handle we consider a special case in which $\theta_{0i}=0$,
i.e. we only have space-space noncommutativity. Since Lorentz
invariance is lost we have to specify the frame in which the above
choice holds. We assume this is the case in the static frame of 
$\pi^0$; namely we work in the static frame of $\pi^0$ in which
$\theta_{0i}=0$. We warn the reader that the following calculation
does not apply to the general case of a moving $\pi^0$ or with
$\theta_{0i}\neq 0$. The unpolarized differential decay rate is
\begin{equation}
\displaystyle d\Gamma=\frac{1}{3!}\frac{1}{2m_{\pi}}~
d\Pi_3~\sum_{\rm pol}|{\cal A}|^2,
\end{equation}
where
\begin{equation}
\begin{array}{rcl} 
\displaystyle\sum_{\rm pol}|{\cal A}|^2&=&
\displaystyle (16eJ)^2\frac{2A^2_3}{d_1d_2d_3}
\left[2^{-4}m^8_{\pi}+(d^4_1+d^4_2+d^4_3)\right],\\
d\Pi_3&=&\displaystyle\Pi_{i=1}^3
\left[\frac{d^3\vec{k}_i}{(2\pi)^32|\vec{k}_i|}
\right](2\pi)^4\delta^4(p-k_1-k_2-k_3).
\end{array}
\end{equation}
The expression for $\displaystyle\sum_{\rm pol}|{\cal A}|^2$ 
simplifies considerably in the current case because we actually
have only one independent angle made of $k_i$ and $\theta_{\mu\nu}$
so that $A_1=A_2=A_3$ and the ${\cal A}_c$ term disappears. 

Special
attention should be paid to the phase space calculation due to the
same reason of Lorentz noninvariance. We might calculate in two steps
as usual; first we integrate over $\vec{k}_1$ and $\vec{k}_2$ in the 
static frame of $\vec{k}_1+\vec{k}_2$, then we move to the static
frame of $\pi^0$ to finish the remaining integration. But this is 
simply wrong without modifying correspondingly the constant matrix 
$\theta_{\mu\nu}$ from one frame to another. Actually, we would 
otherwise obtain a vanishing result for the above specified case.
It turns out that it is much more convenient to fix the frame from the
very beginning and work out kinematics in terms of Euler angles. 
Without loss of generality we assume $\theta_{12}=-\theta_{21}=\theta$
and others vanishing, i.e. the `magnetic' constant field
$B^{i}=\epsilon^{ijk}\theta^{jk}/2$ points in the $z$ direction.
We use the polar and azimuthal angles $\beta$ and $\gamma$ to define
the normal direction of the event plane, and another azimuthal angle
$\alpha$ to fix the absolute direction in the event plane. For example
we may use as a reference direction for this purpose the intersection 
between the event plane and the plane spanned by the normal of the
event plane and the 
$z$ axis although our final result is independent of $\alpha$.
Finishing part of integration using the $\delta^4$ function, we have
\begin{equation}
\displaystyle d\Pi_3=2^{-8}\pi^{-5}~d\omega_1~d\omega_2~d\alpha~
d\cos\beta~d\gamma,
\end{equation}
with $\omega_i=|\vec{k}_i|$. For a given $0\le\omega_1\le m_{\pi}/2$,
we have $(m_{\pi}-2\omega_1)/2\le\omega_2\le m_{\pi}/2$.
The kinematic quantities are computed
below:
\begin{equation}
\begin{array}{rcl}
d_i&=&\displaystyle m_{\pi}(m_{\pi}-2\omega_i)/2,~
\sum_{i=1}^3\omega_i=m_{\pi}/2,\\
\alpha_i&=&\displaystyle \sqrt{m_{\pi}(m_{\pi}-2\omega_1)
(m_{\pi}-2\omega_2)(m_{\pi}-2\omega_3)}(\cos\beta)\theta/4\\
&=&\displaystyle\sqrt{d_1d_2d_3}
\frac{\theta\cos\beta}{\sqrt{2}m_{\pi}}.
\end{array}
\end{equation}
Since $|\theta|m^2_{\pi}\ll 1$, we have
$2A^2_3/(d_1d_2d_3)\approx(\theta m^2_{\pi})^2\cos^2\beta/m^6_{\pi}$.
Completing integration over $\omega_{1,2}$, $\alpha$ and $\gamma$ 
results in the following differential decay rate,
\begin{equation}
\displaystyle\frac{1}{\Gamma}
\frac{d\Gamma}{d\cos\beta}=\frac{3}{2}\cos^2\beta,~~
\displaystyle\Gamma=\frac{\alpha^3 N^2_c m^3_{\pi}}
{2^9~3^3~5\pi^4f^2_{\pi}}(\theta m^2_{\pi})^2.
\end{equation}
Thus, the decay occurs dominantly in the plane which is perpendicular 
to the `magnetic' constant field while it is forbidden in any plane 
which is parallel to the field. This is a particular feature of the 
star product coupling among fields. The simplicity of the above
distribution also has its origin in the much simpler gauge invariant 
structures shown in eq. (19) as compared to the case of the standard 
model. Therefore the differential distributions in the two cases are
completely different.

We note that the $\pi^0\to 2\gamma$ decay is also modified, see
eq. (17). This modification is generally very small and vanishes in
the particular case specified above. So, the branching ratio is,
\begin{equation}
\displaystyle {\rm Br}(\pi^0\to 3\gamma)=
\frac{\Gamma(\pi^0\to 3\gamma)}{\Gamma(\pi^0\to 2\gamma)}=
\frac{\alpha}{120\pi}(\theta m^2_{\pi})^2.
\end{equation}
For a noncommutativity of order $\sqrt{|\theta|}=1~{\rm TeV}^{-1}$, 
we have,
\begin{equation}
\displaystyle {\rm Br}(\pi^0\to 3\gamma)=6.4~10^{-21}.
\end{equation}
This result is larger than an estimate in the standard model by many 
orders of magnitude $\cite{dicus}$, but still far below the current
experimental upper bound of $3.1~10^{-8}$ $\cite{expt}$. 
Further, a much smaller branching ratio will result if some very 
stringent limits on $|\theta|$ are applied.
We also mention in passing that ${\cal L}_{NCA}$ also implies a
C-conserving four photon decay of $\pi^0$. But it is less interesting 
and cannot exceed the standard model process that occurs at one 
loop order in chiral perturbation theory involving anomaly, with a 
branching ratio of order $5.5~10^{-17}$ $\cite{4photon}$.

We have shown how a simple and reasonable generalization of the
anomalous $\pi^0$-photon interaction can lead to the C-violating three
photon decay of the $\pi^0$ in NCQED. We demonstrated explicitly that
for such a consideration to be physically self-consistent it is
mandatory to treat the electrically neutral photon and $\pi^0$ on the
same footing in the sense that they transform under $U(1)$ as if they
were in the adjoint representation of a non-Abelian gauge theory. The
neutral particles thus also enjoy electromagnetic interactions. This
is reminiscent of the wisdom in the usual field theory that one must
keep all possible interactions that are consistent with symmetries 
for the theory to be renormalizable. We are thus inclined to believe
that the above conclusion should be a general result in noncommutative
field theories. Phenomenologically the branching ratio we obtained is
generally much larger than the estimated result in the standard model.
In the case of space-space noncommutativity and with a static $\pi^0$,
the decay distribution has a simple characteristic of being preferred
in a direction specified by the $\theta$ parameter, which would be
quite helpful in expermental identification if there were any chance 
to detect such a small branching ratio.

\vspace{0.5cm}
\noindent
Acknowledgements We are grateful to Klaus Sibold for many encouraging
and helpful discussions and for carefully reading the manuscript.
H.G. enjoyed the stay at ITP, Universit\"at Leipzig where part of work 
was done.

\newpage
\begin{flushleft}
{\Large Figure Captions }
\end{flushleft}
\noindent
Fig. 1 Feynman diagrams for the decay $\pi^0\to 3\gamma$ involving 
the contact interaction ($a$), the NCQED three photon vertex ($b$), 
the normal NCQED $\pi^0$-photon vertex ($c$), and a ghost-anti-ghost 
pair in the final state ($d$). The dashed, wavy and solid lines stand
for the pion, photon and ghost fields respectively. Diagrams obtained
by permutation are not shown.

\newpage
\begin{center}
\begin{picture}(400,600)(0,0)
\SetOffset(10,80)\SetWidth{1.5}

\SetOffset(0,400)
\DashLine(100,110)(100,170){5}
\Photon(100,110)(30,75){3}{7}
\Photon(100,110)(100,75){3}{3.5}
\Photon(100,110)(170,75){-3}{7}
\Text(100,20)[]{\large $(a)$}
\Text(100,180)[]{$p$}
\Text(30,65)[]{$\alpha,k_1$}
\Text(100,65)[]{$\beta,k_2$}
\Text(170,65)[]{$\gamma,k_3$}
\SetOffset(0,-500)

\SetOffset(200,400)
\DashLine(100,110)(100,170){5}
\Photon(100,110)(30,75){3}{7}
\Photon(65,93)(100,75){3}{3.5}
\Photon(100,110)(170,75){-3}{7}
\Text(100,20)[]{\large $(b)$}
\Text(30,65)[]{$\alpha,k_1$}
\Text(100,65)[]{$\beta,k_2$}
\Text(170,65)[]{$\gamma,k_3$}
\SetOffset(-200,-500)

\SetOffset(0,200)
\DashLine(100,110)(100,170){5}
\DashLine(100,110)(65,93){5}
\Photon(65,93)(30,75){3}{3.5}
\Photon(65,93)(100,75){-3}{3.5}
\Photon(100,110)(170,75){-3}{7}
\Text(100,20)[]{\large $(c)$}
\Text(30,65)[]{$\alpha,k_1$}
\Text(100,65)[]{$\beta,k_2$}
\Text(170,65)[]{$\gamma,k_3$}
\SetOffset(0,-200)

\SetOffset(200,200)
\DashLine(100,110)(100,170){5}
\Photon(100,110)(65,93){3}{3.5}
\ArrowLine(30,75)(65,93)
\ArrowLine(65,93)(100,75)
\Photon(100,110)(170,75){-3}{7}
\Text(100,20)[]{\large $(d)$}
\Text(30,65)[]{$k_1$}
\Text(100,65)[]{$k_2$}
\Text(170,65)[]{$\gamma,k_3$}
\SetOffset(-200,-200)

\SetOffset(200,20)
\Text(0,100)[]{\large Figure $1$}
\SetOffset(-200,-20)

\end{picture}\\
\end{center}

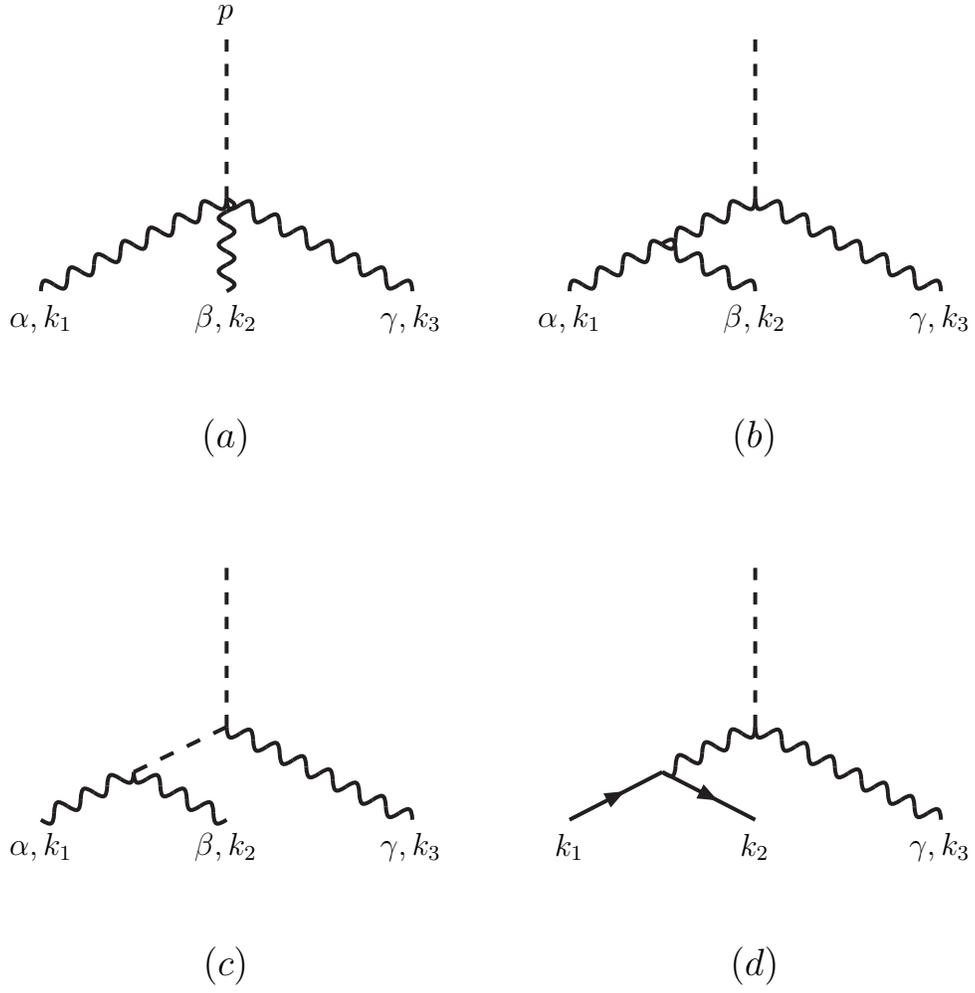
\end{document}